\documentclass[a5paper, 10pt]{article}

\usepackage{cite, amsmath, amssymb, booktabs, lastpage, graphicx}
\usepackage[hidelinks]{hyperref}

\usepackage{geometry}
\usepackage{setspace}
\usepackage{amsthm}
\usepackage{verbatim}

\theoremstyle{plain}

\theoremstyle{remark}

\theoremstyle{definition}

\usepackage{graphicx}
\makeatletter
\renewcommand{\maketitle}{
	\begin{center}
    \rule[.2em]{\textwidth}{0.353mm}
		\begin{minipage}[m]{0.35\textwidth}
			{\scriptsize
				\begin{center}
					\textsf{\textbf{\huge MATCH}\\
						\textit{Communications in Mathematical\\
							and in Computer Chemistry}
					}
			\end{center}}
		\end{minipage}\hfill
		\begin{minipage}[m]{0.65\textwidth}
			\begin{flushright}
				\baselineskip=10px
				{\scriptsize\sffamily{\itshape  MATCH Commun. Math. Comput. Chem.}
				\vol{}
				(\pubyear)
				79--108}\\
				{\scriptsize\sffamily {\bfseries ISSN:} 0340--6253}\\
				{\scriptsize\sffamily \textbf{doi:} \doi{10.46793/match.91-1.079K}}
			\end{flushright}
		\end{minipage}
		\rule[1em]{\textwidth}{.353mm}
		\baselineskip=0.30in
		{\Large\bfseries \@title} \par
		\vspace{5mm}
		\baselineskip=0.2in
		{\large\bfseries \@author}\par
		\vspace{1mm}
		{\it \@address} \par
		{\small\tt \@email} \par
		\vspace{3mm}
		{\small (Received \@date)} \par
	\end{center}
	\vspace{3mm}
}
\newcommand{\address}[1]{\def\@address{#1}}
\newcommand{\email}[1]{\def\@email{#1}}
\newcommand{\doi}[1]{\href{https://doi.org/#1}{#1}}

\newcommand{\vol}{\textbf{91}}
\newcommand{\pubyear}{2024}

\makeatother

\newcommand{\acknowledgment}[1]{\vspace{5mm}\singlespacing
	{\noindent\textbf{\textit{Acknowledgment\/}:} #1}
}

\usepackage{fancyhdr}

\usepackage[margin=1cm,%
			font=footnotesize,%
			format=hang,%
			labelsep=period,%
			labelfont=bf]{caption}

\newtheorem{thm}{Theorem}[section]

\newtheorem{lem}[thm]{Lemma}

\newtheorem{pro}[thm]{Problem}
\newtheorem{exa}[thm]{Example}
\theoremstyle{definition}
\newtheorem{dfn}[thm]{Definition}

\newcommand{\R}{\mathbb R}
\newcommand{\Z}{\mathbb Z}

\newcommand{\de}{\delta}

\newcommand{\ep}{\varepsilon}
\newcommand{\la}{\lambda}

\newcommand{\si}{\sigma}

\newcommand{\ti}{\tilde}

\newcommand{\SM}{\mathrm{SM}}
\newcommand{\LAC}{\mathrm{LAC}}
\newcommand{\PCM}{\mathrm{PCM}}
\newcommand{\PCI}{\mathrm{PCI}}
\newcommand{\WMI}{\mathrm{WMI}}
\newcommand{\RC}{\mathrm{RC}}

\newcommand{\HKS}{\mathrm{HKS}}

\newcommand{\EMD}{\mathrm{EMD}}

\newcommand{\gap}{\mathrm{gap}}

\newcommand{\cov}{\mathrm{Cov}}

\newcommand{\Or}{\mathrm{O}}

\newcommand{\lra}{\leftrightarrow}

\newcommand{\vect}[2]{ \left( \begin{array}{c} #1 \\ #2 \end{array}  \right) }
\newcommand{\mat}[4]{ \left( \begin{array}{cc} 
 #1 & #2 \\ #3 & #4 \end{array} \right)}

\title{Polynomial-Time Algorithms for Continuous Metrics on Atomic Clouds of Unordered Points}
\author{Vitaliy Kurlin$^{a,}$\footnote{Corresponding author.}}

\address{$^a$Materials Innovation Factory and Computer Science department, 
University of Liverpool, Liverpool L69 3BX, United Kingdom  }

\email{vitaliy.kurlin@gmail.com}

\date{July 14, 2023}

\begin{document}

\maketitle

\begin{abstract}
The most fundamental model of a molecule is a cloud of unordered atoms, even without chemical bonds that can depend on thresholds for distances and angles.
The strongest equivalence between clouds of atoms is rigid motion, which is a composition of translations and rotations.
The existing datasets of experimental and simulated molecules require a continuous quantification of similarity in terms of a distance metric.
While clouds of $m$ ordered points were continuously classified by Lagrange's quadratic forms (distance matrices or Gram matrices), their extensions to $m$ unordered points are impractical due to the exponential number of $m!$ permutations.
We propose new metrics that are continuous in general position and are computable in a polynomial time in the number $m$ of unordered points in any Euclidean space of a fixed dimension $n$.
\end{abstract}

\onehalfspacing

\section{Motivations and metric problem statement}
\label{sec:intro}

\begin{figure}[h]
\includegraphics[height=18mm]{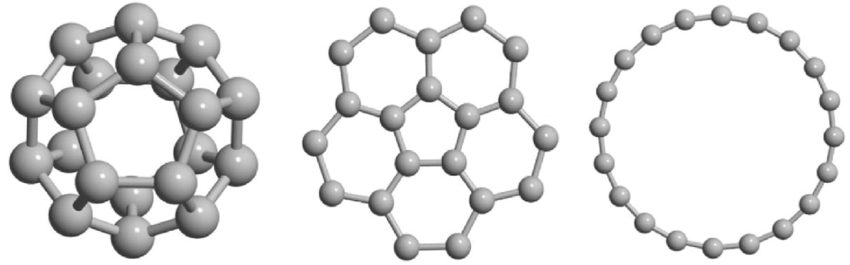}
\hspace*{1mm}
\includegraphics[height=18mm]{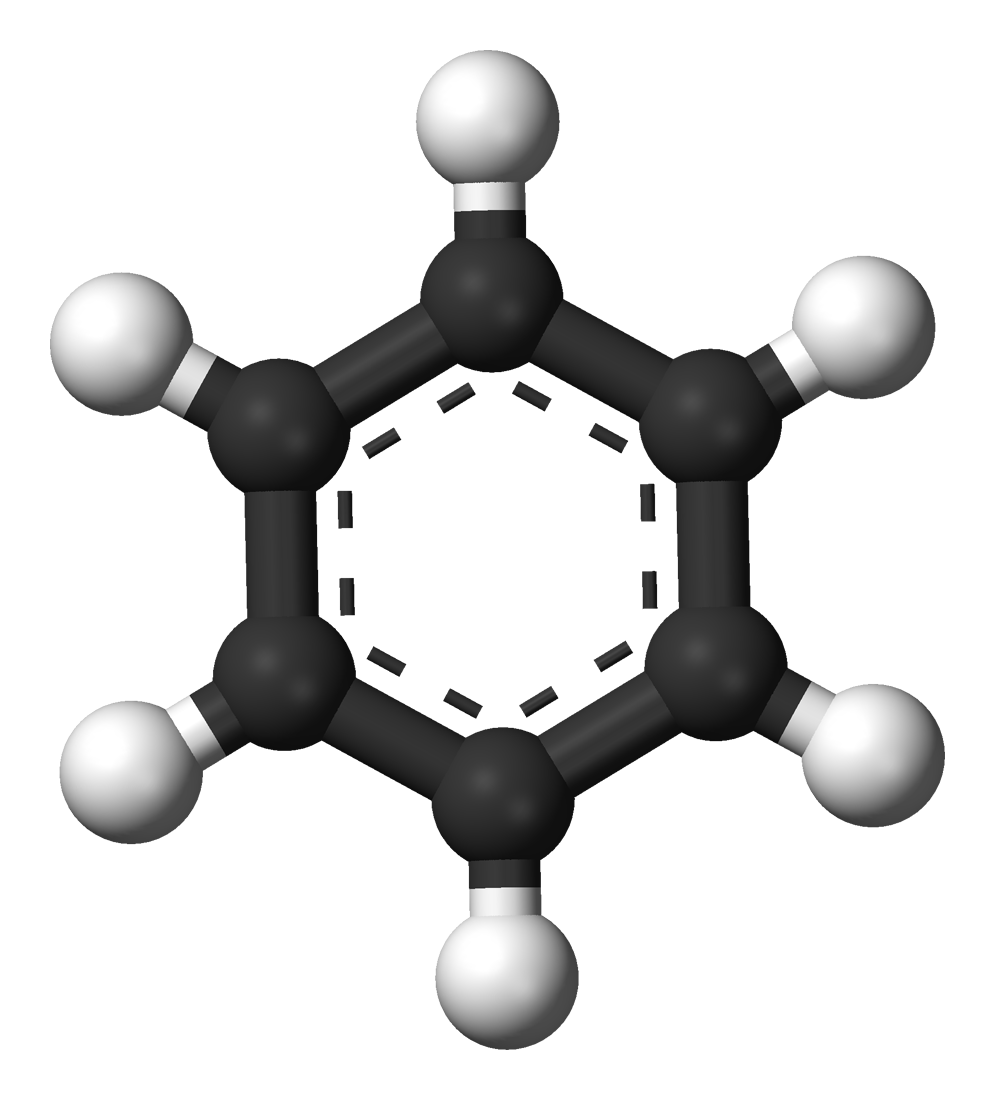}
\hspace*{0.5mm}
\includegraphics[height=18mm]{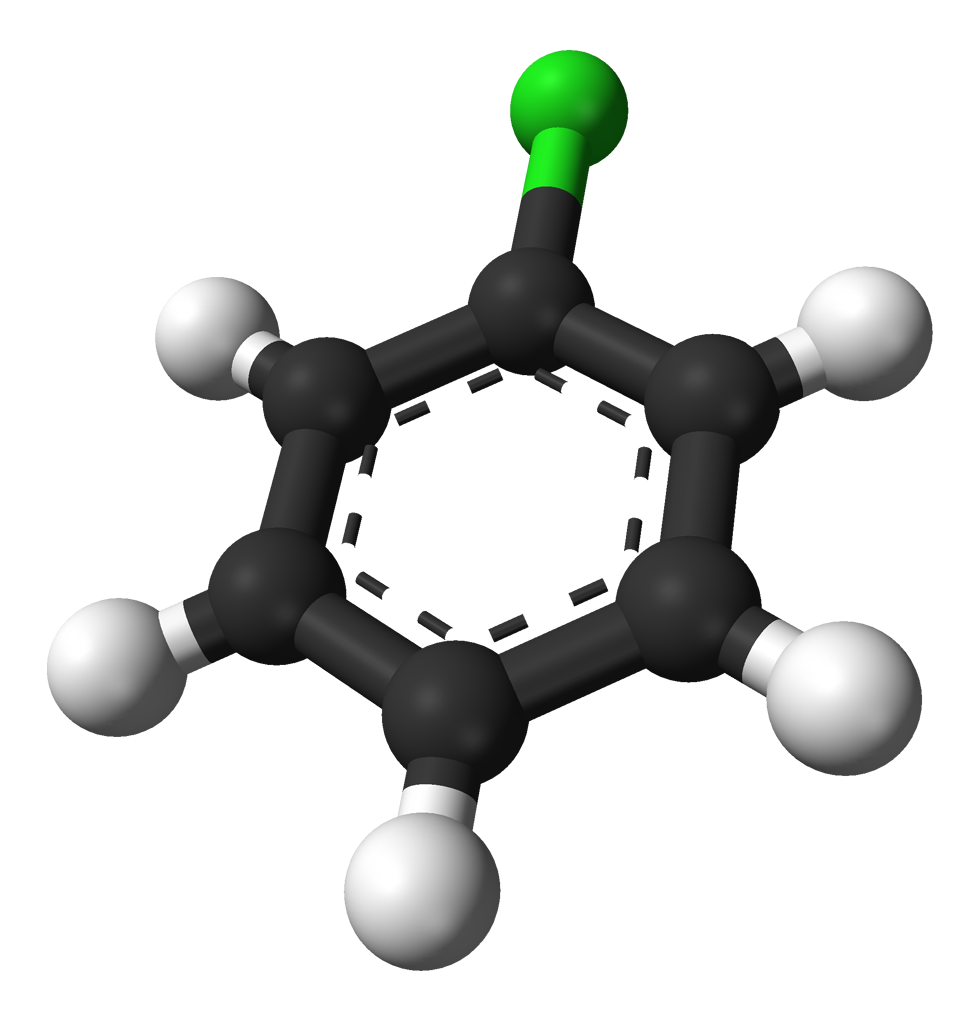}
\hspace*{0.5mm}
\includegraphics[height=18mm]{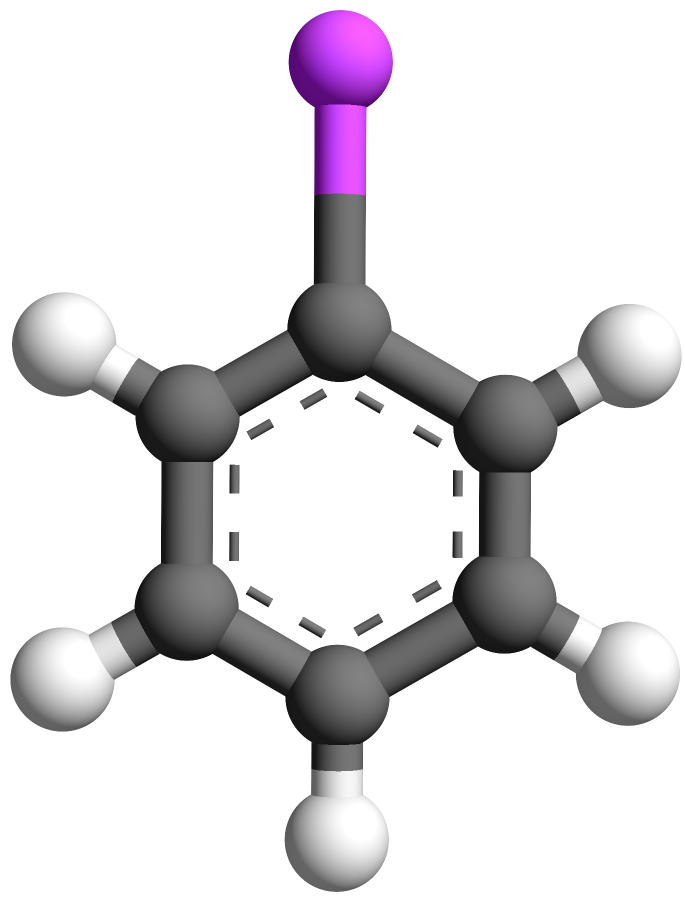}
\caption{
Isomers of $C_{20}$, benzene $C_6 H_6$, phenyllithium $C_6 H_5 Li$, chlorobenzene $C_6 H_5 Li$ have many indistinguishable atoms.
}
\label{fig:molecules}
\end{figure}

Any finite chemical system such as a molecule can be represented as a cloud of atoms whose nuclei are real physical objects \cite{widdowson2022average}, while chemical bonds are not real sticks and only abstractly represent inter-atomic interactions. 
In the hardest scenario, all atoms are modeled as zero-sized points at all atomic centers without any labels such as chemical elements.
For example, the $C_{60}$ molecule \cite{kroto1985c60} consists of 60  unordered carbons.
Allowing different compositions enables a quantitative comparison of isomers, 
see Fig.~\ref{fig:molecules}.
 
Now we formalize the key concepts.
A point \emph{cloud} is any finite set of unordered points in a Euclidean space $\R^n$. 
Since many objects have rigid shapes, the natural equivalence of clouds is a rigid motion or isometry.
\medskip

Any \emph{isometry} of $\R^n$ is a composition of translations, rotations, and reflections represented by matrices from the orthogonal group $O(\R^n)$.
If reflections are excluded, any orientation-preserving isometry $f$ is realized by a \emph{rigid motion} as a continuous family of isometries $f_t:\R^n\to\R^n$, $t\in[0,1]$, where $f_1=f$ and $f_0$ is the identity.
We focus on the isometry because a change of orientation can be easily detected by the sign of the determinant $\det(f(v_1),\dots,f(v_n))$ for a basis $v_1,\dots,v_n$ of $\R^n$.
\medskip

Clouds of unordered points can be decided to be non-isometric only due to an \emph{invariant} \cite{olver1999classical} that is a descriptor preserved under any isometry and all permutations of points.
If points $p_1,\dots,p_m$ are ordered, the matrix of Euclidean distances $|p_i-p_j|$ or the Gram matrix of scalar products $p_i\cdot p_j$ is invariant under isometry \cite{schonemann1966generalized}, but not under $m!$ permutations of points.
\medskip

The exponential number $m!$ of permutations is the major computational obstacle in extending invariants of ordered points to the much harder unordered case. 
Since all atomic coordinates are determined only approximately, all real clouds are not isometric in practice at least slightly.
Hence the important problem is to continuously quantify the difference in terms of a distance metric.
This metric should satisfy all metric axioms, otherwise, the results of clustering algorithms may not be trustworthy \cite{rass2022metricizing}.
\medskip
   
The continuity of a metric in condition (\ref{pro:isometry}d) below is based on 1-1 perturbations of atoms motivated by atomic displacements in real systems.

\begin{pro}[continuous isometry classification of unordered point clouds]
\label{pro:isometry}
Find a complete isometry invariant $I$ and a continuous metric $d$ for any clouds of unordered points in $\R^n$ so that the conditions below hold.
\medskip

\noindent
(\ref{pro:isometry}a) 
\emph{Invariance}: 
if clouds $A\cong B$ are \emph{isometric} in $\R^n$, meaning that $f(A)=B$ for an \emph{isometry} $f:\R^n\to\R^n$, 
then $I(A)=I(B)$, so the invariant $I$ has \emph{no false negatives}, which are pairs $A\cong B$ with $I(A)\neq I(B)$. 
\medskip

\noindent
(\ref{pro:isometry}b) 
\emph{Completeness} :
if $I(A)=I(B)$, then $A\cong B$, so $I$ has \emph{no false positives}, which are pairs of non-isometric $A\not\cong B$ with $I(A)=I(B)$.
\medskip

\noindent
(\ref{pro:isometry}c) 
A \emph{metric} $d$ on invariant values should satisfy all axioms below :  
\smallskip

\noindent
(1) \emph{coincidence} : $d(I(A),I(B))=0$ if and only if $A\cong B$ are isometric; 
\smallskip

\noindent
(2) \emph{symmetry} : $d(I(A),I(B))=d(I(B),I(A))$ for any clouds $A,B\subset\R^n$;
\smallskip

\noindent
(3) \emph{triangle inequality} : $d(I(A),I(C))\leq d(I(A),I(B))+d(I(B),I(C))$.
\medskip

\noindent
(\ref{pro:isometry}d) 
\emph{Continuity}: for $A$ and $\ep>0$, there is $\de$ such that if $B$ is obtained by perturbing points of $A$ in their $\de$-neighborhoods, then $d(I(A),I(B))<\ep$. 
\medskip

\noindent
(\ref{pro:isometry}e) 
\emph{Computability} :
for a fixed $n$, the invariant $I(A)$ and the metric $d(A,B)$ are exactly computable in a polynomial time in the sizes of $A,B$.
\medskip

\noindent
(\ref{pro:isometry}f) 
\emph{Parametrization} : all realizable values $I(A)$ can be parametrized so that any new value of $I$ always gives rise to a reconstructable cloud $A$. 
\end{pro}

In the simplest case of $m=3$ points, all triangles are classified up to isometry (also called congruence in school geometry) by a triple of unordered edge-lengths. 
This Euclid's SSS (side-side-side) theorem was extended to plane polygons 
whose complete invariant is a sequence of edge-lengths considered up to cyclic shifts \cite[Chapter 2, Theorem 1.8]{penner2012decorated}.
\medskip

Section~\ref{sec:PCI} first introduces the Principal Coordinates Invariant (PCI) to classify all clouds that allow a unique alignment by principal directions.
Section~\ref{sec:SM} defines a symmetrized metric on PCIs, which is continuous under perturbations in general position and can be computed (for a fixed dimension $n$) in a subquadratic time in the number of unordered points.
\medskip

Section~\ref{sec:WMI} introduces the Weighted Matrices Invariant (WMI) for any point clouds in $\R^n$.
Section~\ref{sec:LAC+EMD} applies the Linear Assignment Cost and Earth Mover's Distance to define metrics on WMIs, which need only a polynomial time in the number $m$ of points For a fixed dimension $n$.
Section~\ref{sec:discussion} discusses the impact of new results on 
molecular shape recognition.

\section{Past work on point clouds under isometry}
\label{sec:review}

\textbf{The case of ordered points} is much easier than Problem~\ref{pro:isometry}.
Indeed, any ordered points $p_1,\dots,p_m\in\R^n$ can be reconstructed (uniquely up isometry) from the matrix of Euclidean distances $d_{ij}=|p_i-p_j|$ for $i,j=1,\dots,m$ \cite[Theorem~9]{grinberg2019n}.
The equivalent complete invariant is the Gram matrix of scalar products $p_i\cdot p_j$, which can be written and classified in terms of quadratic forms going back to Lagrange in the 18th century.
\medskip

For any clouds $A,B\subset\R^n$ of the same number $m$ of points, the difference between matrices above
can be converted into a continuous metric by taking a matrix norm.
The Procrustes distance between isometry classes of clouds can be computed from the Singular Value Decomposition \cite[appendix~A]{pumir2021generalized}.
All these approaches strongly depend on point order, hence their extensions to unordered points require $m!$ permutations of points.
\medskip

\textbf{Multidimensional scaling} (MDS)
 is a related approach again for a cloud $A$ of $m$ ordered points given by their $m\times m$ distance matrix $D$.
The classical MDS \cite{schoenberg1935remarks} finds an embedding $A\subset\R^k$ (if it exists) preserving all distances of $M$ for a minimum dimension $k\leq m$.
The underlying computation of $m$ eigenvalues of the Gram matrix expressed via $D$ needs $O(m^3)$ time.
The resulting representation of $A\subset\R^k$ uses orthonormal eigenvectors whose ambiguity up to signs for potential comparisons leads to the time factor $2^k$, which can be close to $2^m$.
The new invariant of unordered points needs the much smaller $n\times n$ covariance matrix of a cloud $A\subset\R^n$ and has the faster time $O(n^2m+n^3)$ in Lemma~\ref{lem:PCI_comp}. 
\medskip

\textbf{The crucial difference} between order vs no-order on $m$ points is the exponential number of $m!$ permutations, which are impractical to apply to invariants of ordered points such as distance matrices or Gram matrices.
\medskip

\textbf{Isometry decision} refers to a simpler version of Problem~\ref{pro:isometry} to algorithmically detect a potential isometry between clouds of $m$ unordered points in $\R^n$. 
The algorithm by Brass and Knauer \cite{brass2000testing} takes $O(m^{\lceil n/3\rceil}\log m)$ time, so $O(m\log m)$ in $\R^3$ \cite{brass2004testing}.
The latest advance is the $O(m\log m)$ algorithm in $\R^4$ \cite{kim2016congruence}.
These algorithms output a binary answer (yes/no) without quantifying similarity between clouds by a continuous metric. 
\medskip

\medskip

\textbf{The Hausdorff distance} \cite{hausdorff1919dimension} can be defined for any subsets $A,B$ in an ambient metric space as  $d_H(A,B)=\max\{d_{\vec H}(A,B), d_{\vec H}(B,A) \}$, where the directed Hausdorff distance is 
$d_{\vec H}(A,B)=\sup\limits_{p\in A}\inf\limits_{q\in B}|p-q|$.
To get a metric on rigid shapes, one can further minimize \cite{huttenlocher1993comparing,chew1992improvements,chew1997geometric,chew1999geometric}
 the Hausdorff distance over all isometries $f$ in $\R^n$.
For $n=1$, the Hausdorff distance minimized over translations in $\R$ for sets of at most $m$ points can be found in time $O(m\log m)$ \cite{rote1991computing}.
For $n=2$, the Hausdorff distance minimized over isometries in $\R^2$ for sets of at most $m$ point needs $O(m^5\log m)$ time \cite{chew1997geometric}. 
\medskip

\textbf{Approximate algorithms}.
For a given $\ep>0$ and $n>2$, the related problem to decide if $d_H\leq\ep$ up to translations has the time complexity $O(m^{\lceil(n+1)/2\rceil})$ \cite[Chapter~4, Corollary~6]{wenk2003shape}. 
For general isometry in dimensions $n>2$, approximate algorithms \cite{goodrich1999approximate} tackled minimizations for infinitely many rotations in $\R^3$, later in any $\R^n$ \cite[Lemma~5.5]{anosova2022algorithms}, but the time of exact computations was analyzed only in special cases \cite{marovsevic2018hausdorff,haddad2022hausdorff}. 
\medskip

\textbf{Gromov-Wasserstein distances} are defined between any metric-measure spaces, not necessarily sitting within a common ambient space.
However, even the simplest Gromov-Hausdorff distance for finite metric spaces cannot be approximated within any factor less than 3 in polynomial time unless P=NP \cite[Corollary~3.8]{schmiedl2017computational}.
Gromov-Hausdorff distances were exactly computed for simplices \cite{ivanov2019gromov}, for ultrametric spaces \cite[Algorithm~1]{memoli2021gromov} in $O(m^2)$-time and approximated in polynomial time for metric trees \cite{agarwal2018computing} and in $O(m\log m)$-time for $m$ points in $\R$ \cite[Theorem~3.2]{majhi2023approximating}.
\medskip

\textbf{Topological Data Analysis} studies persistent homology for filtrations of simplicial complexes \cite{carlsson2009topology} on a finite cloud $A$ of unordered points. 
If we consider the standard (Vietoris-Rips, Cech, Delaunay) filtrations, then persistent homology is invariant up to isometry, not up to more general deformations.
Persistence in dimensions 0 and 1 cannot distinguish generic families of inputs \cite{curry2018fiber,catanzaro2020moduli} including non-isometric clouds \cite{smith2022families}.
\medskip

\begin{figure}[h]
\includegraphics[height=16mm]{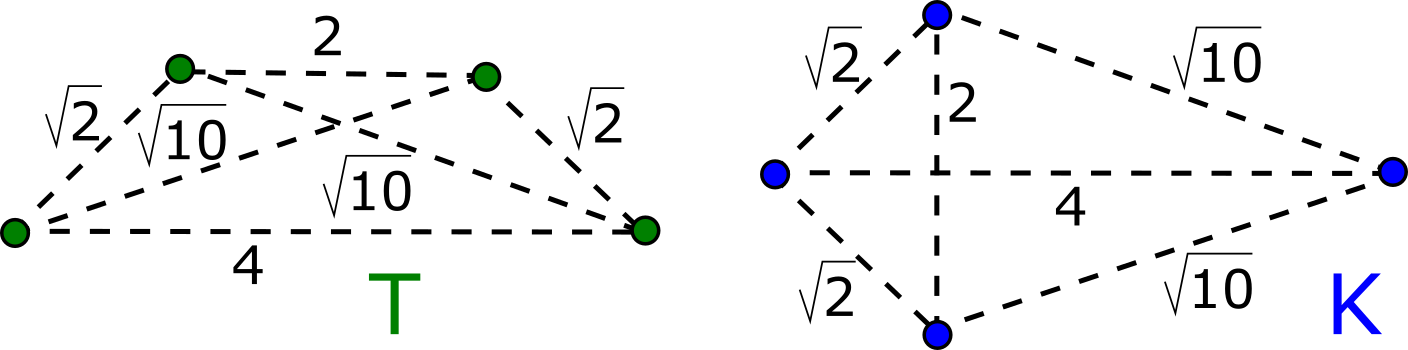}
\hspace*{1mm}
\includegraphics[height=16mm]{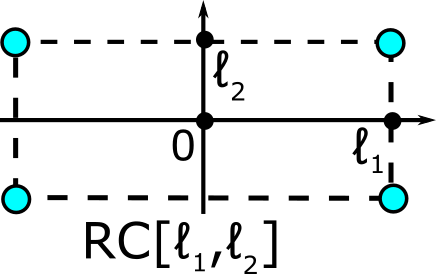}
\hspace*{1mm}
\includegraphics[height=16mm]{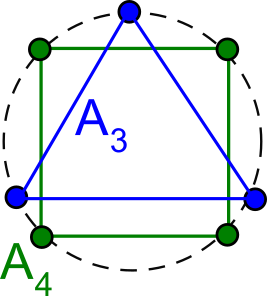}
\caption{
\textbf{First and second}: non-isometric sets $T\not\cong K$ of 4 points have the same 6 pairwise distances.
\textbf{Third}: the vertex set $\RC[l_1,l_2]$ of a $2l_1\times 2l_2$ 
 rectangle. 
\textbf{Fourth}: what is the distance between an equilateral 
triangle $A_3$ and a square $A_4$?
See new invariants and metrics in Examples~\ref{exa:PCI},~\ref{exa:SM},~\ref{exa:triangle-vs-square}.}
\label{fig:4-point_sets}
\end{figure}

\textbf{Distance-based invariants}.
Significant results on matching rigid shapes and registering finite clouds were obtained in \cite{yang2015go, maron2016point,dym2019linearly}.
The total distribution of pairwise distances is complete for point clouds in general position \cite{boutin2004reconstructing}, though infinitely many counter-examples are known, see the non-isometric clouds $T\not\cong K$ of 4 points in the first two pictures of Fig.~\ref{fig:4-point_sets}.

The stronger \emph{local distributions of distances} \cite{memoli2011gromov,memoli2012some}, also known as \emph{shape distributions} \cite{osada2002shape, belongie2002shape, grigorescu2003distance,manay2006integral,pottmann2009integral}
 for metric-measure spaces, are similar to the more specialized \cite{widdowson2022resolving} Pointwise Distance Distributions (PDDs), which can be continuously compared by the Earth Mover's Distance \cite{rubner2000earth}.
\medskip

\textbf{Energy potentials} of molecules use equivariant descriptors of atomic environments \cite{huguenin2023physics}, which are often obtained by deep learning \cite{pozdnyakov2023smooth} and controllably change under rotations.
PDD is conjectured to be complete for finite clouds in $\R^2$ but \cite[Fig.~S4]{pozdnyakov2022incompleteness} provided excellent examples in $\R^3$ that were distinguished only by the stronger invariants in \cite[section~4]{kurlin2023simplexwise}.
\medskip

The latest distance-based invariants \cite{widdowson2023recognizing,kurlin2023strength} satisfy all conditions of Problem~\ref{pro:isometry} apart from parametrization~(\ref{pro:isometry}f).
Indeed, 4 points in the plane have 6 pairwise distances that satisfy one polynomial equation saying that the tetrahedron on these points has volume 0.
Hence randomly sampled 6 positive distances give rise to a real cloud with probability 0.

\section{A complete invariant PCI in a generic case}
\label{sec:PCI}

We start by recalling the Principal Component Analysis (PCA) whose principal directions \cite{abdi2010principal} will be used for building the Principal Coordinates Invariant (PCI).
For any cloud $A\subset\R^n$ of $m$ points has the \emph{center of mass} $\bar A=\dfrac{1}{m}\sum\limits_{p\in A} p$.
Shifting $A$ by the vector $-\bar A$ allows us to always assume that $\bar A$ is the origin $0$. 
Then Problem~\ref{pro:isometry} reduces to invariants only under orthogonal maps from the group $O(n)$ instead of the Euclidean group.

\begin{dfn}[\emph{covariance} matrix $\cov(A)$ of a point cloud $A$]
\label{dfn:covariance}
If we arbitrarily order points $p_1,\dots,p_m$ of a cloud $A\subset\R^n$, we get the sample $n\times m$ matrix (or data table) $P(A)$, whose $i$-th column consists of $n$ coordinates of the point $p_i\in A$, $i=1,\dots,m$.
The \emph{covariance} $n\times n$ matrix $\cov(A)=\dfrac{P(A) P(A)^T}{n-1}$ is symmetric and positive semi-definite meaning that $v^T\cov(A)v\geq 0$ for any vector $v\in\R^n$.
Hence the matrix $\cov(A)$ has real \emph{eigenvalues} $\la_1\geq\dots\geq\la_n\geq 0$ satisfying $\cov(A)v_j=\la_j v_j$ for an \emph{eigenvector} $v_j\in\R^n$, which can be scaled by any real $s\neq 0$.
\end{dfn}

If all eigenvalues of $\cov(A)$ are distinct and positive, there is an orthonormal basis of eigenvectors $v_1,\dots,v_n$ ordered according to the decreasing eigenvalues $\la_1>\cdots>\la_n>0$.
This \emph{eigenbasis} is unique up to reflection $v_j\lra -v_j$ of each eigenvector, $j=1,\dots,n$.

\begin{dfn}[\emph{principally generic} cloud]
\label{dfn:generic}
A point cloud $A\subset\R^n$ is \emph{principally generic} if, after shifting 
$\bar A$ to the origin, the covariance matrix $\cov(A)$ has distinct eigenvalues $\la_1>\cdots>\la_n>0$.
The $j$-th eigenvalue $\la_j$ defines the $j$-th \emph{principal direction} parallel to an eigenvector $v_j$, which is uniquely determined up to scaling.  
\end{dfn}

The vertex set of a rectangle, but not a square, is principally generic.
 
\begin{dfn}[matrix $\PCM$ and invariant $\PCI$]
\label{dfn:PCI}
For $n\geq 1$, let $A\subset\R^n$ be a principally generic cloud of points $p_1,\dots,p_m$ with the center of mass $\bar A$ at the origin $0$ of $\R^n$.
Then $A$ has principal directions along unit length eigenvectors $v_1,\dots,v_n$ well-defined up to a sign.
In the orthonormal basis $V=(v_1,\dots,v_n)^T$, any point $p_i\in A$ has the \emph{principal coordinates} $p_i\cdot v_1,\dots,p_i\cdot v_n$, which can be written as a vertical column $n\times 1$ denoted by $Vp_i$.
The \emph{Principal Coordinates Matrix} is the $n\times m$ matrix $\PCM(A)$ whose $m$ columns are the \emph{coordinate sequences} $Vp_1,\dots,Vp_m$.
Two such matrices are \emph{equivalent} under changing signs of rows due to the ambiguity $v_j\lra -v_j$ of unit length eigenvectors in the basis $V$.
The \emph{Principal Coordinates Invariant} $\PCI(A)$ is an equivalence class of matrices $\PCM(A)$. 
\end{dfn}

For simplicity, we skip the dependence on a basis $V$ in the notation $\PCM(A)$. 
The columns of $\PCM(A)$ are unordered, though we can write them according to any order of points in the cloud $A$ considered as the vector $(p_1,\dots,p_m)$.
Then $\PCM(A)$ can be viewed as the matrix product $VA$ consisting of the $m$ columns $Vp_1,\dots,Vp_m$.

\begin{exa}[computing PCI]
\label{exa:PCI}
\textbf{(a)}
For any $l_1>l_2>0$, let the \emph{rectangular cloud} $\RC[l_1,l_2]$ consist of the four vertices $(\pm l_1,\pm l_2)$ of the rectangle $[-l_1,l_1]\times[-l_2,l_2]$.
Then $\RC[l_1,l_2]$ has the center at $0\in\R^2$ and the sample $2\times 4$ matrix $P=\left( \begin{array}{cccc} 
l_1 & l_1 & -l_1 & -l_1\\
l_2 & -l_2 & l_2 & -l_2
\end{array} \right)$ whose columns are in a 1-1 correspondence with (arbitrarily) ordered points $(l_1,l_2)$, $(l_1,-l_2)$, $(-l_1,l_2)$, $(-l_1,-l_2)$.
The covariance matrix 
$\cov(\RC[l_1,l_2])=\mat{4l_1^2}{0}{0}{4l_2^2}$ has eigenvalues $\la_1=4l_1^2>\la_2=4l_2^2$. 
If we choose unit length eigenvectors $v_1=(1,0)$ and $v_2=(0,1)$, then $\PCM(\RC[l_1,l_2])$ coincides with the matrix $P$ above.
The invariant $\PCI(\RC[l_1,l_2])$ is the equivalence class of all matrices obtained from $P$ by changing signs of rows and re-ordering columns.
\medskip

\noindent
\textbf{(b)}
The vertex set $T$ of the trapezium in the first picture of Fig.~\ref{fig:4-point_sets} has four points written in the columns of
the sample matrix $P(T)=\left( \begin{array}{cccc} 
2 & 1 & -1 & -2 \\
-1/2 & 1/2 & 1/2 & -1/2
\end{array} \right)$ so that the center of mass $\bar T$ is the origin $0$.
Then $\cov(T)=\mat{10}{0}{0}{1}$ has eigenvalues 10, 1 with orthonormal eigenvectors $(1,0)$, $(0,1)$, respectively.
The invariant 
$\PCI(T)$ is the equivalence class of the matrix $P(T)$ above.
The vertex set $K$ of the kite in the second picture of Fig.~\ref{fig:4-point_sets} consists of four points written in the columns of
the sample matrix $P(K)=\left( \begin{array}{cccc} 
5/2 & -1/2 & -1/2 & -3/2 \\
0 & 1 & -1 & 0
\end{array} \right)$ so that the center of mass $\bar K$ is the origin $0$.
Then $\cov(K)=\mat{9}{0}{0}{2}$ has eigenvalues 9, 2 with orthonormal eigenvectors $(1,0),(0,1)$, respectively.
The invariant 
$\PCI(K)$ is the equivalence class of the matrix $P(K)$ above.
\end{exa}

\begin{thm}[generic completeness of $\PCI$]
\label{thm:PCI_completeness}
Any principally generic clouds $A,B\subset\R^n$ of $m$ unordered points 
are isometric if and only if their PCI invariants 
coincide as equivalence classes of matrices. 
\end{thm}
\begin{proof}
Any isometry $f:\R^n\to\R^n$ is a linear map, which maps $A$ to $B$, also sends the center of mass $\bar A$ to the center of mass $\bar B$.
Hence we assume that both centers are at the origin $0\in\R^n$, which is preserved by $f$. 
\smallskip

Any isometry $f$ preserving the origin can be represented by an orthogonal matrix $O_f\in\Or(\R^n)$.
In a fixed orthonormal basis of $\R^n$, let $P(A)$ be the sample matrix of the point cloud $A$.
In the same basis, the point cloud $B$ has the sample matrix $P(B)=O_f P(A)$ and the covariance matrix $\cov(B)=\dfrac{P(B) P(B)^T}{n-1}=\dfrac{O_f (P(A) P(A)^T) O_f^T}{n-1}$.
\smallskip

Any orthogonal matrix $O_f\in\Or(\R^n)$ has the transpose $O_f^T=O_f^{-1}$.
Then $\cov(B)$ is conjugated to $\cov(A)=\dfrac{P(A) P(A)^T}{n-1}$ and has the same eigenvalues as $\cov(A)$, while eigenvectors are related by $O_f$ realizing the change of basis.
If we fix an orthonormal basis of eigenvectors $v_1,\dots,v_n$ for $A$, any point $p\in A$ and its image $f(p)\in B$ have the same coordinates in the bases $v_1,\dots,v_n$ and $f(v_1),\dots,f(v_n)$, respectively.
\smallskip

Hence $\PCM(A),\PCM(B)$ are related by re-ordering of columns (equivalently, points of $A,B$) and by changing signs of rows (equivalently, signs of eigenvectors).
So the equivalence classes coincide: $\PCI(A)=\PCI(B)$.
\smallskip

Conversely, any $n\times m$ matrix $\PCM(A)$ from $\PCI(A)$ contains the coordinates $p_i\cdot v_j$ of points $p_1,\dots,p_m\in A$ in an orthonormal basis $v_1,\dots,v_n$.
Hence all points $p_1,\dots,p_m$ are uniquely determined up to a choice of a basis and isometry of $\R^n$.
\end{proof}

\begin{lem}[time complexity of $\PCI$]
\label{lem:PCI_comp}
For a principally generic cloud $A\subset\R^n$ of $m$ points, a matrix $\PCM(A)$ from the invariant $\PCI(A)$ in Definition~\ref{dfn:PCI} can be computed in time  $O(n^2m+n^3)$.
\end{lem}
\begin{proof}
The computational complexity of finding principal directions \cite{arbenz1992divide} for the symmetric $n\times n$ covariance matrix $\cov(A)$ is $O(n^3)$.
Each of the $nm$ elements of the matrix $\PCM(A)$ can be computed in $O(n)$ time.
Hence the total time is $O(n^2m+n^3)$.
\end{proof}

Theorem~\ref{thm:PCI_completeness} requires that clouds $A,B$ are principally generic, which holds with 100\% probability due to noise.
If real clouds are close to symmetric configurations with equal eigenvalues, to avoid numerical instability, we should use the slower but always complete invariants from section~\ref{sec:WMI}.

\section{A metric on principally generic clouds}
\label{sec:SM}

This section defines a metric on $\PCI$ invariants, whose polynomial-time computation and continuity will be proved in Theorems~\ref{thm:SM_complexity} and~\ref{thm:PCI_continuity}.
For any $v=(x_1,\dots,x_n)\in \R^n$, the \emph{Minkowski norm} is $||v||_{\infty}=\max\limits_{i=1,\dots,n}|x_i|$.
The \emph{Minkowski distance} between $u,v\in\R^n$ is $M_\infty(u,v)=||u-v||_{\infty}$.

\begin{dfn}[bottleneck distance $W_{\infty}$]
\label{dfn:bottleneck}
For clouds $A,B\subset\R^n$ of $m$ points, 
the \emph{bottleneck distance} $W_{\infty}(A,B)=\min\limits_{g:A\to B} \sup\limits_{p\in A}||p-g(p)||_{\infty}$ is minimized over all bijections $g:A\to B$.
\end{dfn}

Below we use the bottleneck distance for a matrix $P$ interpreted as a cloud $[P]$ of its column-vectors in $\R^n$. 

\begin{dfn}[$m$-point cloud ${[P]}\subset\R^n$ of an $n\times m$ matrix $P$]
\label{dfn:matrix-cloud}
For any $n\times m$ matrix $P$, let $[P]$ denote the unordered set of its $m$ columns considered as vectors in $\R^n$.
The set $[P]$ 
can be interpreted as a cloud of $m$ unordered points in $\R^n$.
\end{dfn}

For any $n\times m$ matrices $P,Q$, let $g:[P]\to[Q]$ be a bijection of columns.
Then the Minkowski distance $M_{\infty}(v,g(v))$ between columns $v\in[P]$ and $g(v)\in[Q]$ is the maximum absolute difference of corresponding coordinates in $\R^n$.
The minimization over all column bijections $g:[P]\to[Q]$ gives the bottleneck distance $W_\infty([P],[Q])=\min\limits_{g:[P]\to[Q]}\max\limits_{v\in[P]} M_{\infty}(v,g(v))$ between the sets $[P]$, $[Q]$ considered as clouds of unordered points. 
\medskip

An algorithm for detecting a potential isometry $A\cong B$ will check if $\SM(A,B)=0$ for the metric $\SM$ defined via changes of signs.
A change of signs in $n$ rows can be represented by a binary string $\si$ in the product group $\Z_2^n$, where $\Z_2=\{\pm 1\}$, 1 means no change, $-1$ means a change.
\medskip

For instance, the binary string $\si=(1,-1)\in\Z_2^2$ acts on the matrix $P=\PCM(\RC[l_1,l_2])$ from Example~\ref{exa:PCI} as follows:
$$\si\left( \begin{array}{cccc} 
l_1 & l_1 & -l_1 & -l_1\\
l_2 & -l_2 & l_2 & -l_2
\end{array} \right)
=\left( \begin{array}{cccc} 
l_1 & l_1 & -l_1 & -l_1\\
-l_2 & l_2 & -l_2 & l_2
\end{array} \right).$$

\begin{dfn}[symmetrized metric $\SM$ on matrices and clouds]
\label{dfn:SM}
For any $n\times m$ matrices $P,Q$, the minimization over $2^n$ changes of signs represented by strings $\si\in\Z_2^n$ acting on rows gives the \emph{symmetrized metric} $\SM([P],[Q])
=\min\limits_{\si\in\Z_2^n} W_\infty([\si(P)],[Q])$.
For any principally generic clouds $A,B\subset\R^n$, the \emph{symmetrized metric} is
$\SM(A,B)=\SM([\PCM(A)],[\PCM(B)])$ for any matrices $\PCM(A),\PCM(B)$ from Definition~\ref{dfn:PCI}.
\end{dfn}

If we denote the action of a column permutation $g$ on a matrix $P$ as $g(P)$, the matrix difference $g(P)-Q$ has the Minkowski norm (maximum absolute element) $\max\limits_{v\in[P]} M_{\infty}(v,g(v))$. 
Then $W_\infty([P],[Q])$ will be computed by an efficient algorithm for bottleneck matching in Theorem~\ref{thm:SM_complexity}.

\begin{lem}[metric axioms for the symmetrized metric $\SM$]
\label{lem:SM_axioms}
\textbf{(a)}
The metric $\SM(P,Q)$ from Definition~\ref{dfn:SM} is well-defined on
 equivalence classes of $n\times m$ matrices $P,Q$ considered up to changes of signs of rows and permutations of columns, and satisfies all metric axioms.
\medskip

\noindent
\textbf{(b)}
The metric $\SM(A,B)$ from Definition~\ref{dfn:SM} is well-defined on isometry classes of principally generic clouds $A,B$ and satisfies all axioms.
\end{lem}
\begin{proof}
\textbf{(a)}
The coincidence axiom 
follows from Definition~\ref{dfn:SM}: $\SM([P],[Q])=0$ means that there is a string $\si\in\Z_2^n$ changing signs of rows such that $W_{\infty}([\si(P)],[Q])=0$.
By the coincidence axiom for $W_{\infty}$, the point clouds $[\si(P)],[Q]\subset\R^n$ should coincide, hence $[Q]$ is obtained from $[P]$ by a compositions of reflections in the axes $x_i$ with $\si_i=-1$.
The symmetry follows due to inversibility of $\si\in\Z_2^n$ and the symmetry of 
$W_\infty$, so
$\SM([P],[Q])=\min\limits_{\si\in\Z_2^n} W_\infty([\si(P)],[Q])=\min\limits_{\si^{-1}\in\Z_2^n} W_\infty([P],[\si^{-1}(Q)])=\SM([Q],[P]).$

To prove the triangle inequality $\SM(P,M)+\SM(Q,M)\geq\SM(P,Q)$, let binary strings $\si_P,\si_Q\in\Z_2^n$ be optimal for $\SM(P,M)$ and $\SM(Q,M)$, respectively, in Definition~\ref{dfn:SM}.
The triangle inequality for 
$W_\infty$ 
implies that 
\begin{equation*}
  \begin{split}
\SM(P,M)+\SM(Q,M) & =W_\infty([\si_P(P)],[M])
+W_\infty([\si_Q(Q)],[M]) \\ & \geq W_\infty([\si_P(P)],[\si_Q(Q)]).
  \end{split}
\end{equation*}

Since applying the same change $\si_Q^{-1}$ of signs in both matrices $\si_P(P)$ and $\si_Q(Q)$ does not affect the minimization for all changes of signs, the final expression equals $W_\infty([\si_Q^{-1}\circ\si_P(P)],[Q])$ and has the lower bound $\SM(P,Q)=\min\limits_{\si\in\Z_2^n} W_\infty([\si(P)],[Q])$ due to the minimization over all $\si\in\Z_2^n$ instead of one string $\si_Q^{-1}\circ\si_P$ in $\Z_2^n$.
\medskip

\noindent
\textbf{(b)}
The coincidence axiom follows from Theorem~\ref{thm:PCI_completeness}: $A\cong B$ are isometric if and only if $\PCI(A)=\PCI(B)$ meaning that any matrices $\PCM(A),\PCM(B)$ representing the equivalence classes $\PCI(A),\PCI(B)$, respectively, become identical after a column permutation $g:[\PCM(A)]\to[\PCM(B)]$ and the change of signs of rows by a binary string $\si\in\Z_2^n$.
Indeed, $M_{\infty}(v,g(v))=0$ for all columns $v$ in the matrix $\si(\PCM(A))$ means that the matrices $\si(\PCM(A))$ and $\PCM(B)$ become identical after the column permutation $g$.
The symmetry and triangle axioms for $\SM(A,B)$ follow from part \textbf{(a)} for the matrices $P=\PCM(A)$ and $Q=\PCM(B)$.
\end{proof}

\begin{exa}[computing the symmetrized metric $\SM$]
\label{exa:SM}
\textbf{(a)}
By Example~\ref{exa:PCI}(a), the vertex set $\RC[l_1,l_2]$ of any rectangle with sides $2l_1>2l_2$ in the plane has $\PCI$ represented by the matrix 
$\PCM(\RC[l_1,l_2])=\left( \begin{array}{cccc} 
l_1 & l_1 & -l_1 & -l_1\\
l_2 & -l_2 & l_2 & -l_2
\end{array} \right)$.
The vertex set $\RC[l_1',l_2']$ of any other rectangle has a similar matrix whose element-wise subtraction from $\PCM(\RC[l_1,l_2])$ consists of $\pm l_1\pm l_1'$ and $\pm l_2\pm l_2'$.
Re-ordering columns and changing signs of rows minimizes the maximum absolute value of these elements to $\max\{|l_1-l_1'|,|l_2-l_2'|\}$, which should equal $\SM(\RC[l_1,l_2],\RC[l'_1,l'_2])$.
\medskip

\noindent
\textbf{(b)}
The invariants $\PCI$ of the vertex sets $T$ and $K$ in Fig.~\ref{fig:4-point_sets} were computed in Example~\ref{exa:PCI}(b) and represented by these matrices from Definition~\ref{dfn:PCI}:
\begin{equation*}
  \begin{split}
\PCM(T) & =
\left( \begin{array}{cccc} 
2 & 1 & -1 & -2 \\
-1/2 & 1/2 & 1/2 & -1/2
\end{array} \right), \\
\PCM(K) &=
\left( \begin{array}{cccc} 
5/2 & -1/2 & -1/2 & -3/2 \\
0 & 1 & -1 & 0
\end{array} \right).
  \end{split}
\end{equation*}

The maximum absolute value of the element-wise difference of these matrices is $|1-(-\frac{1}{2})|=\frac{3}{2}$, which cannot be smaller after permuting columns and changing signs of rows.
The symmetrized metric equals $\SM(T,K)=W_{\infty}(\PCM(T),\PCM(K))=\frac{3}{2}$.
\end{exa}

\begin{thm}[time of the metric $\SM$]
\label{thm:SM_complexity}
\textbf{(a)}
Given any $n\times m$ matrices $P,Q$, the symmetrized metric $\SM(P,Q)$ in Definition~\ref{dfn:SM} is computable in time $O(m^{1.5}2^n\log^{n}m)$.
If $n=2$, the time is $O(m^{1.5}\log m)$.
\medskip

\noindent
\textbf{(b)}
The above conclusions hold for $\SM(A,B)$ 
of any principally generic $m$-point clouds $A,B\subset\R^n$ represented by $n\times m$ matrices $\PCM(A),\PCM(B)$. 
\end{thm}
\begin{proof}
\textbf{(a)}
For a fixed binary string $\si\in\Z_2^n$, \cite[Theorem~6.5]{efrat2001geometry} computes the bottleneck distance $W_{\infty}(\si(P),Q)$ between the clouds $[P],[Q]$ of $m$ points in time $O(m^{1.5}\log^n m)$ with space $O(m\log^{n-2} m)$.
If $n=2$, the time is $O(m^{1.5}\log m)$ by \cite[Theorem~5.10]{efrat2001geometry}.
The minimization for all binary strings $\si\in\Z_2^n$ brings the extra factor $2^n$.
\medskip

\noindent
\textbf{(b)}
It follows from part \textbf{(a)} for $P=\PCM(A)$ and $Q=\PCM(B)$.
\end{proof}

Lemmas~\ref{lem:matrix_norms} and~\ref{lem:eigenvectors} will help prove the continuity of the symmetrizied metric $\SM$ under perturbations 
in Theorem~\ref{thm:PCI_continuity}.
Recall that any $n\times n$ matrix $E$ has the \emph{2-norm} $||E||_2=\sup\limits_{|v|=1}|E v|$ and the maximum norm
$||E||_{\infty}=\max\limits_{j=1,\dots,n}\sum\limits_{k=1}^n|E_{jk}|$.
If the center of mass $\bar A=0\in\R^n$ is the origin, define the \emph{radius} $r_A=\max\limits_{p\in A}|p|$. 

\begin{lem}[upper bounds for matrix norms]
\label{lem:matrix_norms}
Let $A,B\subset\R^n$ be any principally generic clouds of $m$ points with covariance matrices $\cov(A)$ and $\cov(B)$, respectively.
Set $u=\dfrac{nm}{n-1} W_{\infty}(A,B) (r_A+r_B)$.
Then
\begin{equation}
||\cov(A)-\cov(B)||_2\leq u \text{ and }
||\cov(A)-\cov(B)||_{\infty}\leq u.
\end{equation}
\end{lem}
\begin{proof}
Assume that $A,B$ have centers of mass at the origin 0.
Let $g:A\to B$ be a bijection minimizing the bottleneck distance $W_{\infty}(A,B)$.
Let $A$ consist of $m$ points $p_1,\dots,p_m$.
Set $\ti p_i=g(p_i)$ for $i=1,\dots,m$.
Let $x_j(p)$ denote the $j$-th coordinate of a point $p\in\R^n$, $j=1,\dots,n$.
The covariance matrices can be expressed as follows:
\begin{equation*}
\cov(A)_{jk} = \dfrac{1}{n-1}\sum\limits_{i=1}^m x_j(p_i) x_k(p_i), \quad
\cov(B)_{jk} = \dfrac{1}{n-1}\sum\limits_{i=1}^m x_j(\ti p_i) x_k(\ti p_i).
\end{equation*}

Since the Minkowski distance $M_{\infty}(p_i,\ti p_i)\leq W_{\infty}(A,B)$, the upper bounds $|x_j(p_i)-x_j(\ti p_i)|\leq W_{\infty}(A,B)$ hold for all $i=1,\dots,m$ and $j=1,\dots,n$, and will be used below to estimate each element of the $n\times n$ matrix $E=\cov(A)-\cov(B)$ as follows: 
$(n-1)|E_{jk}|=$ 
\begin{equation*}\begin{split}
& \leq\sum\limits_{i=1}^m |x_j(p_i) x_k(p_i)-x_j(\ti p_i) x_k(\ti p_i)| \\
& =\sum\limits_{i=1}^m \Big|x_j(p_i)\big(x_k(p_i)-x_k(\ti p_i)\big) 
+x_k(\ti p_i)\big(x_j(p_i)-x_j(\ti p_i)\big)\Big| \\
& \leq  \sum\limits_{i=1}^m \Big( |x_j(p_i)|\cdot \big|x_k(p_i)-x_k(\ti p_i)\big|
+ |x_k(\ti p_i)|\cdot \big|(x_j(p_i)-x_j(\ti p_i)\big| \Big) \\
& \leq W_{\infty}(A,B)\sum\limits_{i=1}^m \Big(|x_j(p_i)|+ |x_k(\ti p_i)|\Big)\leq m W_{\infty}(A,B) (r_A+r_B).
\end{split}\end{equation*}

If we denote the final expression by $w$, the required bound is $u=\dfrac{nw}{n-1}$.
Let $E_1,\dots,E_n\in\R^n$ be the rows of $E=\cov(A)-\cov(B)$.
Then $||E||_2\leq$
\begin{equation*}\begin{split}
 & =\sup\limits_{|v|=1}|E v|
\leq \sup\limits_{|v|=1} \sqrt{\sum\limits_{j=1}^n (E_j\cdot v)^2}
\leq \sup\limits_{|v|=1} \sqrt{\sum\limits_{j=1}^n |E_j|^2|v|^2} 
\leq \sqrt{\sum\limits_{j=1}^n |E_j|^2} \\ & 
=\sqrt{\sum\limits_{j,k=1,\dots,n}E_{jk}^2}
\leq \sqrt{n^2\max\limits_{j,k=1,\dots,n}E_{jk}^2}
=n\max\limits_{j,k=1,\dots,n}|E_{jk}|
\leq \dfrac{nw}{n-1}=u.
\end{split}\end{equation*}
 
Finally, $||E||_{\infty}=\max\limits_{j=1,\dots,n}\sum\limits_{k=1}^n|E_{jk}|\leq \dfrac{nw}{n-1}=u$ as required.
\end{proof}

The result below is quoted in a simplified form for the PCA case.

\begin{lem}[eigenvector perturbation {\cite[Theorem~3]{fan2018eigenvector}}]
\label{lem:eigenvectors}
Let $C$ be a symmetric $n\times n$ matrix whose eigenvalues $\la_1>\dots>\la_n>0$ have a minimum $\gap(C)=\min\limits_{j=1,\dots,n}(\la_j-\la_{j+1})>0$, where $\la_{n+1}=0$.
Let $v_i,\tilde v_i$ be unit length eigenvectors of $C$ and its symmetric perturbation $\tilde C$ such that $E=C-\tilde C$ has the 2-norm $||E||_2<\gap(C)/2$.
Then $\max\limits_{j=1,\dots,n}|v_j-\tilde v_j|=O\left(\dfrac{n^{3.5}\mu^2||E||_{\infty}+n\sqrt{\mu}||E||_2}{\gap(C)} \right)$, where the \emph{incoherence} $\mu$ is the maximum sum of squared $j$-th coordinates of $v_1,\dots,v_n$ for $j=1,\dots,n$, which has the rough upper bound $n$.
\end{lem}

\begin{thm}[continuity of $\SM$]
\label{thm:PCI_continuity}
For any principally generic cloud $A\subset\R^n$ 
and any $\ep>0$, there is $\de>0$ (depending on $A$ and $\ep$) such that
if any principally generic cloud $B\subset\R^n$ has $W_{\infty}(A,B)<\de$, then $\SM(A,B)<\ep$.
\end{thm}
\begin{proof}
Let $g:A\to B$ be a bijection minimizing the distance $W_{\infty}(A,B)$ so that 
$M_{\infty}(p,g(p))=W_{\infty}(A,B)$
for $p\in A,g(p)\in B$. 
By Lemma~\ref{lem:matrix_norms} the difference 
$E=\cov(A)-\cov(B)$ has the matrix norms bounded by $u=\dfrac{nm}{n-1} W_{\infty}(A,B) (r_A+r_B)$.
By Lemma~\ref{lem:eigenvectors} with $\mu\leq n$ the maximum difference of eigenvectors of $C=\cov(A)$ and $\cov(B)$ has the 
norm 
\begin{equation*}
\max\limits_{j=1,\dots,n}|v_j-\tilde v_j|
\leq O\left(\dfrac{n^{5.5}w}{\gap(C)}\right)
=O\left(\dfrac{n^{5.5}}{\gap(C)}\right)
m W_{\infty}(A,B) \big(r_A+r_B\big).
\end{equation*}

The bijection $g:A\to B$ induces a bijection between the columns of the matrices $\PCM(A),\PCM(B)$ so that the column represented by any point $p_i\in A$ maps to the column represented by $\ti p_i=g(p_i)\in B$.
We can permute the columns of $\PCM(B)$ so that the columns represented by $p_i,\ti p_i$ have the same index $i$.
Let $v_1,\dots,v_n$ and $\ti v_1,\dots,\ti v_n$ be unit length eigenvectors of $\cov(A),\cov(B)$, respectively.
Then we estimate 
\begin{equation*}\begin{split}
& |p_i\cdot v_j-\ti p_i\cdot \ti v_j|=|(p_i-\ti p_i)\cdot v_j+\ti p_i\cdot (v_j-\ti v_j)|  \\ &
\leq |p_i-\ti p_i|\cdot |v_j|+|\ti p_i|\cdot |v_j-\ti v_j|
\leq |p_i-\ti p_i|+r_B \max\limits_{j=1,\dots,n}|v_j-\tilde v_j|.
\end{split}\end{equation*}
The final maximum satisfies 
$\max\limits_{j=1,\dots,n}|v_j-\tilde v_j|\leq (r_A+r_B) O\Big(\dfrac{n^{5.5}m}{\gap(C)}\Big)$, where $C=\cov(A)$.
Since $r_B\leq r_A+W_{\infty}(A,B)$, we get the following upper bound for element-wise difference $\PCM(A)-\PCM(B)$.
\begin{equation*}\begin{split}
& |p_i\cdot v_j-\ti p_i\cdot \ti v_j|
\leq W_{\infty}(A,B)\left(1+r_B(r_A+r_B) O\Big(\dfrac{n^{5.5}m}{\gap(C)}\Big)\right) \\ &
\leq W_{\infty}(A,B)\left(1+\big(r_A+W_{\infty}(A,B)\big)\big(2r_A+W_{\infty}(A,B)\big)O\Big(\dfrac{n^{5.5}m}{\gap(C)}\Big)\right).
\end{split}\end{equation*}
For any $\ep>0$, one can choose $\de>0$ (depending only on $A$, not on $B$) so that if $W_{\infty}(A,B)<\de$ then
$|p_i\cdot v_j-\ti p_i\cdot \ti v_j|<\ep$ for any $i=1,\dots,m$ and $j=1,\dots,n$.
Then the $i$-th columns $u_i\in[\PCM(A)]$ and $u'_i\in[\PCM(B)]$ have the Minkowski distance $M_{\infty}(u_i,u'_i)<\ep$ for all $i=1,\dots,m$.
Hence $\SM(A,B)<\ep$ by Definition~\ref{dfn:SM}, as required for the continuity.
\end{proof}

\section{The complete invariant WMI for all clouds}
\label{sec:WMI}

This section extends the invariant PCI from Definition~\ref{dfn:PCI} to a complete invariant WMI (Weighted Matrices Invariant) of all possible clouds. 
\smallskip

If a cloud $A\subset\R^n$ is not principally generic, some of the eigenvalues $\la_1\geq\dots\geq\la_n\geq 0$ of the covariance matrix $\cov(A)$ coincide or vanish.
Let us start with the most singular case when all eigenvalues are equal to $\la>0$.
The case $\la=0$ means that $A$ is a single point. 
Though $A$ has no preferred (principal) directions, $A$ still has the well-defined center of mass $\bar A=\dfrac{1}{m}\sum\limits_{p\in A}p$, which is at the origin $0\in\R^n$ as always.
For $n=2$, we consider $m$ possible vectors from the origin $0$ to every point of $A-\{0\}$.

\begin{dfn}[Weighted Matrices Invariant $\WMI(A)$ for clouds  $A\subset\R^2$]
\label{dfn:WMI2}
Let a cloud $A$ of $m$ points $p_1,\dots,p_m$ in $\R^2$ have the center of mass at the origin $0$.
For any point $p_i\in A-\{0\}$, let $v_1$ be the unit length vector parallel to $p_i\neq 0$.
Let $v_2$ be the unit length vector orthogonal to $v_1$ whose anti-clockwise angle from $v_1$ to $v_2$ is $+\dfrac{\pi}{2}$.
The $2\times m$ matrix $M(p_i)$ consists of the $m$ pairs of coordinates of all points $p\in A$ written in the orthonormal basis $v_1,v_2$, for example, $p_i=\vect{||p_i||_2}{0}$. 
Each matrix $M(p_i)$ is considered up to re-ordering of columns.
If one point $p$ of $A$ is the origin $0$, there is no basis defined by $p=0$, let $M(p)$ be the zero matrix in this centered case.
If $k>1$ of the matrices $M(p_i)$ are \emph{equivalent} up to re-ordering of columns, we collapse them into one matrix with the weight $\dfrac{k}{m}$.
The unordered collection of the equivalence classes of $M(p)$
with weights for all $p\in A$ is called the \emph{Weighted Matrices Invariant} $\WMI(A)$.
\end{dfn}

In comparison with the generic case in Definition~\ref{dfn:PCI}, for any fixed $i=1,\dots,m$, if $p_i\neq 0$, then the orthonormal basis $v_1,v_2$ is uniquely defined without the ambiguity of signs, which will re-emerge for higher dimensions $n>2$ in Definition~\ref{dfn:WMI} later.

\begin{exa}[regular clouds $A_m\subset\R^2$]
\label{exa:reg_clouds}
Let $A_m$ be the vertex set of a regular $m$-sided polygon inscribed into a circle of a radius $r$, see the last picture in Fig.~\ref{fig:4-point_sets}.
Due to the $m$-fold rotational symmetry of $A_m$, the invariant $\WMI(A_m)$ consists of a single matrix (with weight 1) whose columns
 are the vectors $\vect{r\cos\frac{2\pi i}{m}}{r\sin\frac{2\pi i}{m}}$, $i=1,\dots,m$.
For instance, the vertex set $A_3$ of the equilateral triangle has $\WMI(A_3)=
\left\{\left( \begin{array}{ccc} 
r & -r/2 & -r/2 \\
0 & r\sqrt{3}/2 & -r\sqrt{3}/2
\end{array} \right)\right\}$.
The vertex set $A_4$ of the square has $\WMI(A_4)=
\left\{\left( \begin{array}{cccc} 
r & 0 & 0 & -r \\
0 & r & -r & 0
\end{array} \right)\right\}$.
Let  $B_m$ be obtained from $A_m$ by adding the origin $0\in\R^2$.
Then $\WMI(B_m)$ has the matrix from $\WMI(A_m)$ with the weight $\dfrac{m}{m+1}$ and the zero $2\times 4$ matrix with the weight $\dfrac{1}{m+1}$ representing the added origin $0$.
\end{exa}

Definition~\ref{dfn:WMI} applies to all point clouds $A\subset\R^n$ including the most singular case when all eigenvalues of the covariance matrix $\cov(A)$ are equal, so we have no preferred directions at all.

\begin{dfn}[Weighted Matrices Invariant for any cloud $A\subset\R^n$]
\label{dfn:WMI}
Let a cloud $A\subset\R^n$ of $m$ points $p_1,\dots,p_m$ have the center of mass at the origin $0$.
For any ordered sequence of points $p_1,\dots,p_{n-1}\in A$, build an orthonormal basis $v_1,\dots,v_n$ as follows.
The first unit length vector $v_1$ is $p_1$ normalized by its length.
For $j=2,\dots,n-1$, the unit length vector 
$v_j$ is $p_j-\sum\limits_{k=1}^{j-1}(p_j\cdot v_k)v_k$ normalized by its length.
Then every $v_j$ is orthogonal to all previous vectors $v_1,\dots,v_{j-1}$ and belongs to the $j$-dimensional subspace spanned by $p_1,\dots,p_j$.
Define the last unit length vector $v_n$ by its orthogonality to $v_1,\dots,v_{n-1}$ and the positive sign of the determinant $\det(v_1,\dots,v_n)$ of the matrix with the columns $v_1,\dots,v_n$.
\smallskip

The $n\times m$ matrix $M(p_1,\dots,p_{n-1})$ consists of column vectors of all points $p\in A$ in the basis $v_1,\dots,v_n$, for example, $p_1=(||p_1||_2,0,\dots,0)^T$. 
If $p_1,\dots,p_{n-1}\in A$ are affinely dependent, let $M(p_1,\dots,p_{n-1})$ be the $n\times m$ matrix of zeros in this centered case.
If $k>1$ matrices are \emph{equivalent} up to re-ordering of columns, we collapse them into a single matrix with the weight $\dfrac{k}{N}$, where $N=m(m-1)\dots(m-n+1)$.
The \emph{Weighted Matrices Invariant} $\WMI(A)$ is the unordered set of equivalence classes of matrices $M(p_1,\dots,p_{n-1})$ with weights for all sequences of points $p_1,\dots,p_{n-1}\in A$.
\end{dfn}

If $\cov(A)$ has some equal eigenvalues, $\WMI(A)$ can be made smaller by choosing bases only for subspaces of eigenvectors with the same eigenvalue.

\begin{thm}[completeness of $\WMI$]
\label{thm:WMI_completeness}
\textbf{(a)}
Any clouds $A,B\subset\R^n$ are related by rigid motion (orientation-preserving isometry) if and only if 
there is a bijection $\WMI(A)\to\WMI(B)$ preserving all weights or, equivalently,
some matrices $P\in\WMI(A)$, $Q\in\WMI(B)$ are related by re-ordering of columns.
So $\WMI(A)$ is a complete invariant of $A$ up to 
rigid motion.
\medskip

\noindent
\textbf{(b)}
Any mirror reflection $f:A\to B$ induces a bijection $\WMI(A)\to\WMI(B)$ respecting their weights and changing the sign of the last row of every matrix.
This pair of $\WMI$s is a complete invariant of $A$ up to isometry including reflections.
\end{thm}
\begin{proof}
\textbf{(a)}
As in the proof of Theorem~\ref{thm:PCI_completeness}, let the centers $\bar A,\bar B$ coincide with the origin $0\in\R^n$.
Given an orientation-preserving isometry $f:\R^n\to\R^n$ mapping $A$ to $B$, any ordered sequence $p_1,\dots,p_{n-1}\in A$ maps to $f(p_1),\dots,f(p_{n-1})\in B$.
Since $f$ is a linear map preserving all scalar products and lengths of vectors, we conclude that 
$$f(p_j-\sum\limits_{k=1}^{j-1}(p_j\cdot v_k)v_k)=f(p_j)-\sum\limits_{k=1}^{j-1}(f(p_j)\cdot f(v_k))f(v_k).$$ 
By Definition~\ref{dfn:WMI} the isometry $f$ maps the orthonormal basis $v_1,\dots,v_n$ of the sequence $p_1,\dots,p_{n-1}\in A$ to the orthonormal basis $f(v_1),\dots,f(v_n)$ of the sequence $f(p_1),\dots,f(p_{n-1})\in B$. 
Then any point $p\in A$ has the same coordinates $p\cdot v_j=f(p)\cdot f(v_j)$, $j=1,\dots,n$, in the basis $v_1,\dots,v_n$ as its image $f(p)\in B$ in the basis $f(v_1),\dots,f(v_n)$.
The matrices $M(p_1,\dots,p_{n-1})\in\WMI(A)$ and $M(f(p_1),\dots,f(p_{n-1}))\in\WMI(B)$ coincide if their columns (equivalently, points of $A,B$) are matched by 
$f$.
\smallskip

By choosing any 
$p_1,\dots,p_n\in A$, the isometry $f:A\to B$ induces the bijection $\WMI(A)\to\WMI(B)$ respecting the weights of matrices (equivalent up to re-ordering of columns).
So condition \textbf{(a)} holds and implies \textbf{(b)} saying that some $P\in\WMI(A)$ and $Q\in\WMI(B)$ are equivalent.
\smallskip

Conversely, if a matrix $P\in\WMI(A)$ coincides with $Q\in\WMI(B)$, let $v_1,\dots,v_n$ and $u_1,\dots,u_n$ be the orthonormal bases used for writing these matrices in Definition~\ref{dfn:WMI}.
The isometry $f$ mapping $v_1,\dots,v_n$ to $u_1,\dots,u_n$ maps $A$ to $B$ because any point $p\in A$ in the basis $v_1,\dots,v_n$ has the same coordinates as its image $f(p)\in B$ in the basis $f(v_1),\dots,f(v_n)$.
\medskip

\noindent
\textbf{(b)}
Let $f:\R^n\to\R^n$ be any orientation-reversing isometry such as a mirror reflection.
For any sequence of affinely independent points $p_1,\dots,p_{n-1}\in A$, the matrix $M_A(p_1,\dots,p_{n-1})$ from Definition~\ref{dfn:WMI} describes $A$ in the basis defined by $p_1,\dots,p_{n-1}$ with a fixed orientation of $\R^n$.
\smallskip

Composing $f$ with a rigid motion 
moving $f(p_1),...,f(p_{n-1})$ back to $p_1,\dots,p_{n-1}$, respectively, we can assume that $f$ fixes each of $p_1,\dots,p_{n-1}$, while $\WMI$ is preserved by part \textbf{(a)}.
Then $f$ is the mirror reflection $A\to B$ in the hyperspace spanned by the fixed points $p_1,\dots,p_{n-1}$.
Since the basis vector $v_n$ is uniquely defined by $p_1,\dots,p_{n-1}$ for a fixed orientation of $\R^n$, any other point $p\in A$ maps to its mirror image $f(p)\in B$, so $p$ and $f(p)$ have opposite projections to $v_n$.
Then the matrix $M_B(p_1,\dots,p_{n-1})$ describing $f(A)=B$ in the basis $v_1,\dots,v_n$ differs from $M_A(p_1,\dots,p_{n-1})$ by the
change of sign in the last row.
\smallskip

Hence $f$ induces a bijection $\WMI(A)\to\WMI(B)$, where each matrix changes the sign of its last row and is considered up to permutation of columns.
Conversely, any matrix from $\WMI(A)$ whose last row is considered up to a change of sign suffices to reconstruct $A$ up to
isometry.
\end{proof}

One can store in computer memory only one matrix $M(p_1,\dots,p_{n-1})$ from the full $\WMI(A)$ whose elements parametrize the isometry class of $A$ as required by (\ref{pro:isometry}f).
Any such matrix suffices to reconstruct a point cloud $A$ up to orientation-preserving isometry of $\R^n$ by Theorem~\ref{thm:PCI_completeness}.
The full invariant $\WMI(A)$ can be computed from the reconstructed cloud. 

\begin{lem}[time of $\WMI$]
\label{lem:WMI_complexity}
For any cloud $A\subset\R^n$ of $m$ points and any sequence $p_1,\dots,p_{n-1}\in A$, the matrix $M(p_1,\dots,p_{n-1})$ from Definition~\ref{dfn:WMI} can be computed in time $O(nm+n^3)$.
All $N=m(m-1)\dots(m-n+1)=O(m^{n-1})$ matrices in the Weighted Matrices Invariant $\WMI(A)$ can be computed in time $O((nm+n^3)N)=O(nm^n+n^3m^{n-1})$.  
\end{lem}
\begin{proof}
For a fixed sequence $p_1,\dots,p_{n-1}\in A$, the vectors $v_1,\dots,v_{n-1}$ are computed by 
Definition~\ref{dfn:WMI} in time $O(n^2)$.
The last vector $v_n$ might need the $O(n^3)$ computation of 
$\det(v_1,\dots,v_n)$.
Every point $p\in A$ can be re-written in this basis as $p=\sum\limits_{j=1}^n (p\cdot v_j)v_j$ in time $O(n)$.
Hence the matrix $M(p_1,\dots,p_{n-1})$ is computed in time $O(nm+n^3)$.
Since there are exactly $N=m(m-1)\dots(m-n+1)$ ordered sequences of points $p_1,\dots,p_{n-1}\in A$, all matrices in $\WMI(A)$ are computed in time $O((nm+n^3)N)$.
\end{proof}

\section{Exactly computable metrics 
all clouds}
\label{sec:LAC+EMD}

This section introduces two metrics on Weighted Matrices Invariants ($\WMI$s), which are computable in polynomial time by Theorems~\ref{thm:LAC_time} and~\ref{thm:EMD_time}.
Since any isometry $f:A\to B$ induces a bijection $\WMI(A)\to\WMI(B)$, we will use a linear assignment cost \cite{jonker1987shortest} based on permutations of matrices.

\begin{dfn}[Linear Assignment Cost LAC]
\label{dfn:LAC}
Recall that Definition~\ref{dfn:SM} introduced the bottleneck distance $W_{\infty}$ on matrices considered up to re-ordering of columns.
For any clouds $A,B\subset\R^n$ of $m$ points,
consider the \emph{Linear Assignment Cost}  
$\LAC(A,B)=\min\limits_{g}\sum\limits_{P\in\WMI(A)} W_{\infty}(P,g(P))$ minimized \cite{jonker1987shortest} over all bijections $g:\WMI(A)\to\WMI(B)$ of full Weighted Matrices Invariants consisting of all $N=m(m-1)\dots(m-n+1)$ equivalence classes of matrices.
\end{dfn}

\begin{lem}[$\LAC$ on clouds]
\label{lem:LAC}
\textbf{(a)}
The Linear Assignment Cost from Definition~\ref{dfn:LAC} satisfies all metric axioms on 
clouds under rigid motion. 
\medskip

\noindent
\textbf{(b)}
Let $A'$ be any mirror image of $A$.
Then $\min\{\LAC(A,B),\LAC(A',B)\}$ is a metric on classes of clouds up to general isometry including reflections.
\end{lem}
\begin{proof}
\textbf{(a)}
The only non-trivial coincidence axiom follows from Theorem~\ref{thm:WMI_completeness} and the coincidence axiom of the bottleneck distance $W_{\infty}$: any clouds $A,B$ are isometric if and only if there is a bijection $\WMI(A)\to\WMI(B)$ matching all matrices up to permutations of columns, so all corresponding matrices have bottleneck distance $W_{\infty}=0$. 
\medskip

\noindent
\textbf{(b)}
All axioms for 
follow from the relevant axioms for $\LAC(A,B)$.
\end{proof}

\begin{thm}[time complexity of $\LAC$ on $\WMI$s]
\label{thm:LAC_time}
For any clouds $A,B\subset\R^n$ of $m$ points, the invariants $\WMI(A),\WMI(B)$ consists of at most $N=m(m-1)\dots(m-n+1)=O(m^{n-1})$ matrices.
Then the metric $\LAC(A,B)$ from Definition~\ref{dfn:LAC} 
can be computed in time $O(m^{1.5}(\log^n m)N^2+N^3)=O(m^{2n-0.5}\log^n m+m^{3n-3})$.
If $n=2$, the time is $O(m^{3.5}\log m)$.
\end{thm}
\begin{proof}
By \cite[Theorem~6.5]{efrat2001geometry}, for any matrices $P\in\WMI(A)$ and $Q\in\WMI(B)$, the bottleneck distance $W_{\infty}([P],[Q])$ can be computed in time $O(m^{1.5}\log^n m)$.
For $N\times N$ pairs of such matrices, computing all costs $c(P,Q)=W_{\infty}([P],[Q])$ takes
$O(m^{1.5}(\log^n m) N^2)$ time.
If $n=2$, \cite[Theorem~5.10]{efrat2001geometry} reduces 
the time of all costs $W_{\infty}([P],[Q])$ to $O(m^{1.5}(\log m) N^2)$.
Using the same time factor $O(N^2)$, one can check if $c(P,Q)=0$, which means that the clouds $[P]\cong[Q]$ are isometric.
Finally, with all $N^2$ costs $c(P,Q)$ ready, the algorithm by Jonker and Volgenant \cite{jonker1987shortest} computes the Linear Assignment Cost $\LAC(A,B)$ in the extra time $O(N^3)$.
\end{proof}

The worst-case estimate $N=O(m^{n-1})$ of the size (number of matrices in) $\WMI(A)$ is very rough.
If the covariance matrix $\cov(A)$ has equal eigenvalues, $\WMI(A)$ is often smaller due to extra symmetries of $A$.
\medskip

However, for $n=2$, even the rough estimate of the LAC time $O(m^{3.5}\log m)$ improves the time $O(m^5\log m)$ for computing the exact Hausdorff distance between $m$-point clouds under Euclidean motion in $\R^2$.
\medskip

Since real noise may include erroneous points, it is practically important to continuously quantify the similarity between close clouds consisting of different numbers of points.
The weights of matrices allow us to match them more flexibly via the Earth Mover's Distance \cite{rubner2000earth} than via strict bijections $\WMI(A)\to\WMI(B)$. 
The Weighted Matrices Invariant $\WMI(A)$ can be considered as a finite distribution $C=\{C_1,\dots,C_k\}$ of matrices (equivalent up to re-ordering columns) with weights.

\begin{dfn}[Earth Mover's Distance on weighted distributions]
\label{dfn:EMD}
Let $C=\{C_1,\dots,C_k\}$ be a finite unordered set of objects with weights $w(C_i)$, $i=1,\dots,k$.
Consider another set $D=\{D_1,\dots,D_l\}$ with weights $w(D_j)$, $j=1,\dots,l$.
Assume that a distance between any objects $C_i,D_j$ is measured by a metric $d(C_i,D_j)$.
A \emph{flow} from $C$ to $D$ is a $k\times l$ matrix  whose entry $f_{ij}\in[0,1]$ represents a partial \emph{flow} from an object $C_i$ to $D_j$.
The \emph{Earth Mover's Distance} is the minimum \emph{cost} 
$\EMD(C,D)=\sum\limits_{i=1}^{k} \sum\limits_{j=1}^{l} f_{ij} d(C_i,D_j)$ over $f_{ij}\in[0,1]$ subject to 
$\sum\limits_{j=1}^{l} f_{ij}\leq w(C_i)$ for $i=1,\dots,k$, 
$\sum\limits_{i=1}^{k} f_{ij}\leq w(D_j)$ for $j=1,\dots,l$, and
$\sum\limits_{i=1}^{k}\sum\limits_{j=1}^{l} f_{ij}=1$.
\end{dfn}

The first condition $\sum\limits_{j=1}^{l} f_{ij}\leq w(C_i)$ means that not more than the weight $w(C_i)$ of the object $C_i$ `flows' into all objects $D_j$ via the flows $f_{ij}$, $j=1,\dots,l$. 
Similarly, the second condition $\sum\limits_{i=1}^{k} f_{ij}\leq w(D_j)$ means that all $f_{ij}$ from $C_i$ for $i=1,\dots,k$ `flow' into $D_j$ up to its weight $w(D_j)$.
\medskip

The last condition
$\sum\limits_{i=1}^{k}\sum\limits_{j=1}^{l} f_{ij}=1$
 forces to `flow' all $C_i$ to all $D_j$.  
The EMD is a partial case of more general Wasserstein metrics \cite{vaserstein1969markov} in transportation theory \cite{kantorovich1960mathematical}.
For finite distributions as in Definition~\ref{dfn:EMD}, the metric axioms for $\EMD$ were proved in \cite[appendix]{rubner2000earth}. 
$\EMD$ can compare any weighted distributions of different sizes.
Instead of the bottleneck distance $W_{\infty}$ on columns on $\PCM$ matrices, one can consider $\EMD$ on the distributions of columns (with equal weights) in these matrices.

\begin{lem}[time complexity of EMD on distributions of columns]
\label{lem:EMD_matrices}
Any matrix $P$ of a size $n\times m(P)$ can be considered as a distribution of $m(P)$ columns with equal weights $\frac{1}{m(P)}$.
For two such matrices $P,Q$ having the same number $n$ of rows but potentially different numbers $m(P),m(Q)$ of columns, measure the distance between any columns by the Minkowski metric $M_{\infty}$ in $\R^n$.
For the matrices $P,Q$ considered as weighted distributions of columns, the Earth Mover's Distance $\EMD(P,Q)$ can be computed in time $O(m^3\log m)$, where $m=\max\{m(P),m(Q)\}$.
\end{lem}
\begin{proof}
EMD needs $O(m^3\log m)$ time \cite{goldberg1987solving} for distributions of size $m$.
\end{proof}

\begin{thm}[time of $\EMD$ on clouds]
\label{thm:EMD_time}
Let clouds $A,B\subset\R^n$ of up to $m$ points have pre-computed invariants $\WMI(A),\WMI(B)$ of sizes at most $N\leq m(m-1)\dots(m-n+1)=O(m^{n-1})$.
Measure the distance between any matrices $P\in\WMI(A)$ and $Q\in\WMI(B)$ as 
$\EMD(P,Q)$ from Lemma~\ref{lem:EMD_matrices}.
Then  the Earth Mover's Distance $\EMD(\WMI(A),\WMI(B))$ 
from Definition~\ref{dfn:EMD} 
can be computed in time $O(m^3(\log m) N^2+N^3\log N)=O((m^{2n+1} +nm^{3n-3})\log m)$.
\end{thm}
\begin{proof}
By Lemma~\ref{lem:EMD_matrices}, the metric $\EMD(P,Q)$ can be computed in time $O(m^3\log m)$.
For $N\times N$ pairs of such matrices, computing all costs $c(P,Q)=\EMD(P,Q)$ takes $O(m^3(\log m) N^2)$ time.
With all costs ready, $\EMD(A,B)$ is computed \cite{goldberg1987solving}
 in the extra time $O(N^3\log N)$.
\end{proof}

\begin{exa}[$\EMD$ for a square and equilaterial triangle]
\label{exa:triangle-vs-square}
Let $A_4$ and $A_3$ be the vertex sets of a square and equilateral triangle inscribed into the circle of a radius $r$ in Example~\ref{exa:reg_clouds}.
$\PCM(A_3)=
\left( \begin{array}{ccc} 
r & -r/2 & -r/2 \\
0 & r\sqrt{3}/2 & -r\sqrt{3}/2
\end{array} \right)$ and
$\PCM(A_4)=
\left( \begin{array}{cccc} 
r & 0 & 0 & -r \\
0 & r & -r & 0
\end{array} \right)$.
Notice that switching the signs of the 2nd row keeps the PCI matrices the same up to permutation of columns.
The weights of the three columns in $\PCM(A_3)$ are $\dfrac{1}{3}$.
The weights of the four columns in $\PCM(A_4)$ are $\dfrac{1}{4}$.
The EMD optimally matches the identical first columns of $\PCM(A_3)$ and $\PCM(A_4)$ with weight $\dfrac{1}{4}$ contributing the cost $0$.
The remaining weight $\dfrac{1}{3}-\dfrac{1}{4}=\dfrac{1}{12}$ of the first column $\vect{r}{0}$ in $\PCM(A_3)$ can be equally distributed between the closest (in the $M_{\infty}$ distance) columns $\vect{0}{\pm r}$ contributing the cost $\dfrac{r}{12}$.
The column $\vect{-r}{0}$ in $\PCM(A_4)$ has equal distances $M_{\infty }=\dfrac{r}{2}$ to the last columns $\vect{-r/2}{\pm r\sqrt{3}/2}$ in $\PCM(A_3)$ contributing the cost $\dfrac{r}{8}$.
Finally, the distance $M_{\infty}=\dfrac{r}{2}$ between the columns $\vect{0}{\pm r}$ and $\vect{-r/2}{\pm r\sqrt{3}/2}$ with the common signs is counted with the weight $\dfrac{5}{24}$ and contributes the cost $\dfrac{5r}{48}$.
The final optimal flow $(f_{jk})$ matrix 
$\left( \begin{array}{cccc} 
1/4 & 1/24 & 1/24 & 0 \\
0 & 5/24 & 0 & 1/8 \\
0 & 0 & 5/24 & 1/8
\end{array} \right)$
gives $\EMD(\PCM(A_3),\PCM(A_4))=\dfrac{r}{12}+\dfrac{r}{8}+\dfrac{5r}{48}=\dfrac{5r}{16}$.
\end{exa}

\section{Discussion of significance for atomic clouds}
\label{sec:discussion}

Problem~\ref{pro:isometry} was stated in the hard case for clouds of unordered points in any $\R^n$ because real shapes such as atomic clouds from molecules and salient points from laser scans often include indistinguishable points.
This paper complements many past advances 
by rigorous proofs for all singular point clouds whose principal directions are undefined in $\R^n$.
\medskip

The Principal Coordinates Invariant (PCI) should suffice for object retrieval \cite{rubner2000earth,sun2009concise} and other applications in Computer Vision and Graphics, because real clouds are often principally generic due to noise in measurements.
Then, for any fixed dimension $n$, Theorem~\ref{thm:SM_complexity} computes the symmetrized metric $\SM$ on PCIs faster than in a quadratic time in the number $m$ of points.
The key insight was the realization that Principal Component Analysis (PCA) belongs not only to classical statistics but also provides easily computable metrics for point clouds under isometry.
Though sensitivity of PCA under noise was studied for years, Theorem~\ref{thm:PCI_continuity} required more work and recent advances to guarantee the continuity of PCI.  
\medskip


The Weighted Matrices Invariant (WMI) completely parametrizes the moduli space of $m$-point clouds under isometry.
The complete classification in Theorem~\ref{thm:WMI_completeness} goes far beyond the state-of-the-art parametrizations, which are available for moduli spaces of point clouds only in dimension~2 \cite{penner2012decorated}.
For proteins and other molecules in $\R^3$, the moduli space was described only under continuous deformations not respecting distances \cite{penner2016moduli}.
However, non-isometric embeddings of the same protein can have different physical and chemical properties such as binding to drug molecules, and hence should be continuously distinguished by computable metrics. 
\smallskip

This paper focused on foundations, so experiments are postponed to future work.
The exactly computable metric on WMIs can be adapted to the complete isometry invariants of periodic crystals \cite{anosova2021isometry}, which has been done only for 1-periodic sequences  \cite{anosova2022density,anosova2023density,kurlin2022computable}.
The earlier invariants \cite{widdowson2022average, widdowson2021pointwise, widdowson2022resolving} detected geometric duplicates, which had wrong atomic types but were deposited in the well-curated (mostly by experienced eyes) world's largest collection of real materials (Cambridge Structural Database).
Another problem is to prove the continuity of WMIs under perturbations of clouds whose subsets are linearly independent.
Though the complexity in Theorem~\ref{thm:SM_complexity} is practical in dimensions $n=2,3$, it is still important to improve the complexity of the symmetrized metric $\SM$ for higher dimensions. 

\acknowledgment{
This research was supported by the Royal Academy Engineering Fellowship IF2122/186,  EPSRC New Horizons EP/X018474/1, and Royal Society APEX fellowship APX/R1/231152.
The author thanks all members of the Data Science Theory and Applications group at the University of Liverpool 
and all reviewers for their time and suggestions.
}

\singlespacing

\bibliographystyle{match}       
\bibliography{metrics-atomic-clouds}

\end{document}